# Hawking Radiation: A Comparison of Pure-state and Thermal Descriptions

王 一帆 Yi-Fan Wang

Masterarbeit in Physik
angefertigt im Institut für Theoretische Physik
der Universität zu Köln

vorgelegt der
Mathematisch-Naturwissenschaftlichen Fakultät
der
Rheinischen Friedrich-Wilhelms-Universität
Bonn

February 2017

I hereby declare that this thesis was formulated by myself and that no sources or tools other than those cited were used.

Bonn, ................                    ..................................
          Date                                                    Signature

2. Gutachter:   Prof. Dr. Hans-Peter Nilles
1. Gutachter:   Prof. Dr. Claus Kiefer

To 劉 青藍 (Qinglan Liu) I dedicate this thesis, for all her love and support.

# Contents

# Introduction

> Our mistake is not that we take our theories too seriously, but that we do not take them seriously enough.
>
> Steven Weinberg [1]

Although widely accepted, Hawking radiation remains a field of heated discussion and active research, especially in the interpretation and extrapolation of it. Being one of the first conclusions of the original calculation, Hawking temperature was deduced from a *pure state* of the quantum field in the background space-time of a collapsing body, whereas a temperature in statistical physics is usually derived from a statistical ensemble in equilibrium, described by a *thermal* and *mixed state*. This discrepancy itself has since long been largely ignored, albeit related issues have always been in spotlight, for instance the *information loss problem*, the *origin of black hole entropy* [2, 3], etc. A detailed investigation of the pure and thermal descriptions would help understanding the aforementioned questions by laying them on a more solid foundation.

In this work, which is motivated by [4, 5], the focus is to reveal and quantify the difference between the two cases mentioned above. Chapter 2 is a review of Hawking radiation in the (3 + 1)-dimensional Einstein gravitation which is basically along Hawking's original line, as well as that in (1 + 1)-dimensional dilaton gravity, also known as the Callan–Giddings–Harvey–Strominger (CGHS) model, in which the quantum field theory in curved space-time not only can be *derived* from the full theory of quantum gravity, but also be solved exactly. Then in chapter 3, motivated by the coincidence of the particle-number expectations, which are the diagonal elements of the density operators, the author computes the correlator of field strength, so as to reveal the difference in the off-diagonal elements. In chapter 4, inspired by a new foundation of statistical physics, which is based on exact results in quantum information theory, the author quantifies the discrepancy between the two cases by calculating the trace distance between them, and the results are shown to be in accordance with the quantum-informational foundation, as well as those in chapter 3. Chapter 5 is summary and outlook.

In the appendices, appendix A explains more details about the CGHS model, appendix B collects some useful results in quantised simple harmonic oscillator, and appendix C introduces trace distance and fidelity.

Throughout the text, the *natural units* will be used unless specified, where the speed of light in vacuum $c$, the reduced Planck constant $\hbar$ and Boltzmann constant $k$ are all set to unity, while the Newton constant $G$ is kept.

# Hawking Radiation

## 2.1 $(3+1)$-dimensional Einstein gravitation

Based on the advances in the geometry of space-time and 'mechanics' of black holes, as well as in quantum field theory in curved space-time, Hawking quantitatively revealed a semi-classical property of the black holes in [7]. In his model, the Einsteinian gravitational background is fixed to be that of a spherically collapsing body[1], the conformal diagram of which is shown in fig. 2.1. A neutral (real), massless scalar field $\phi(x)$ is minimally coupled to the gravitational background, so the action of the model reads

$$S := \int_{\mathcal{M}} \mathrm{d}^4 x \sqrt{-g} \left\{ g^{\mu\nu}(\partial_\mu \phi)(\partial_\nu \phi) \right\}. \tag{2.1.1}$$

One quantises the scalar field canonically on the Cauchy surface $\mathcal{I}^-$ by introducing ladder operators $a^\mp$'s, and on $\mathcal{I}^+ \cup \mathcal{H}^+$ by $b^\mp$'s (for $\mathcal{I}^+$) and $c^\mp$'s (for $\mathcal{H}^+$). An *early-time vacuum* is defined by

$$a^-(p) |h\rangle := 0, \tag{2.1.2}$$

where $a^-(p)$ annihilates a particle with momentum $p$, so that

$$\langle n_a(p) \rangle_h := \langle h | n_a(p) | h \rangle \equiv 0, \qquad \forall p, \tag{2.1.3}$$

where $n_a(p) = a^+(p) a^-(p)$ is the number operator of mode with momentum $p$. This means that an asymptotic observer at early time, whose definition of particles agrees with $a^\mp$, detects no particle.

For asymptotic observers, Hawking was able to evaluate the *late-time* properties, or those with respect to the $b$'s, of $|h\rangle$. He derived in [9] that[2]

$$\langle n_b(\omega) \rangle_h \approx \Gamma_\omega \left( \mathrm{e}^{2\pi\omega/\kappa} - 1 \right)^{-1}, \tag{2.1.4}$$

where $0 < \Gamma_\omega < 1$ is the *grey-body factor*, and $\omega$ being the angular frequency of the mode regarded. Comparing eq. (2.1.4) with that of a Bose–Einstein distribution, or a *black-body* radiation field,

$$\langle n(\omega) \rangle_{\mathrm{BE}} = \left( \mathrm{e}^{\omega/T} - 1 \right)^{-1}, \tag{2.1.5}$$

it is appealing to conclude that eq. (2.1.4) describes a *grey-body* radiation field, with the tem-

---

[1] A detailed account for space-times of collapsing body can be found in [8].
[2] Possible divergent $\delta(0)$ factors in the particle-number expectations have been systematically ignored.

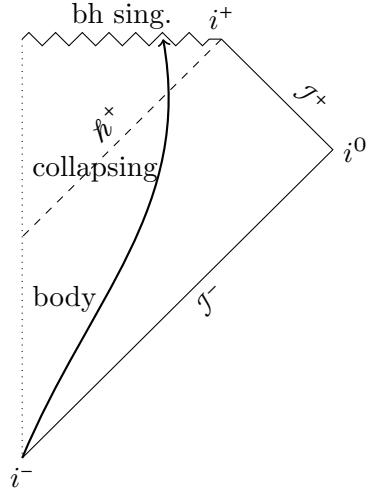

Figure 2.1: A schematic conformal diagram of a spherically collapsing body in Einstein gravitation, in which massive matter (presumably a star) spherically collapses by gravitational interaction. The boundary of the collapsing body is denoted by the thick line with arrow. The quantum field is solved on $\mathcal{J}^-$ and $\mathcal{J}^+ \cup \hbar^+$.

perature
$$T_{\mathrm{H}} := \kappa/2\pi \equiv \hbar/ck \cdot \kappa/2\pi, \qquad (2.1.6)$$
also named after Hawking, and an absorption coefficient $\Gamma_\omega$.

Hawking himself argued that his calculation also holds for a matter field with spin, as well as for the collapse result being a rotating black hole. Further quantities could also be obtained in the background of an eternal black hole, whose connection to an astrophysical collapsing body has been constructed, showing the evaluation is physically robust [10]. The algebraic approach to quantum fields showed that it is also mathematically reliable [11].

Though widely acknowledged, Hawking temperature is in fact technically ill-defined, if we take the calculation *seriously enough*. Recall that a temperature $T$ of a quantum system can only be defined if the system is in a thermal equilibrium, which can be described by a density operator
$$\rho = Z^{-1} e^{-H/T} \qquad (2.1.7)$$
of a *mixed state*, where $H$ is the Hamiltonian of the system, and $Z = \operatorname{tr} e^{-H/T}$ is the partition function. For a bosonic system, the density operator reads
$$\rho_{\mathrm{BE}} \sim Z^{-1} \sum_E e^{-E/T} |E\rangle \langle E|, \qquad (2.1.8)$$
where $(E, |E\rangle)$ are the energy eigenvalue and the corresponding eigenstate of the system, respectively. As $|h\rangle$ is a *pure state*, whose density operator reads
$$\rho_h := |h\rangle \langle h|, \qquad (2.1.9)$$
it is drastically different from eq. (2.1.8). One naturally asks, how different are the pure and mixed states, and what does it imply?

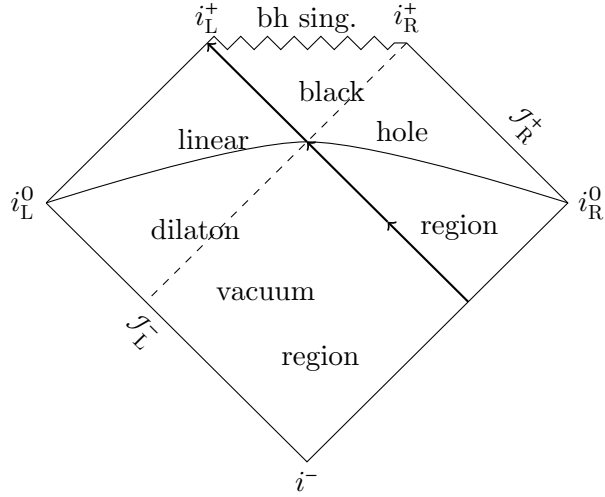

Figure 2.2: The conformal diagram of collapsing null shell in (1 + 1) dimensional dilaton gravity. The black-hole horizon is marked by the dashed line, whereas the ingoing null shell is expressed by the thick line with arrow. The *linear dilaton vacuum region* and the *black hole region* mark the space-time blocks before and after the ingoing shell, respectively. The quantum field, on the other hand, is solved on the hyper-surface connecting $i^0_{L/R}$ and the intersection of the null shell and the horizon.

## 2.2 In $(1 + 1)$-dimensional dilaton gravity

In the Einsteinian background of a collapsing body, in which Hawking did his calculation, no analytic solution of a quantum field theory has so far been found due to technical difficulties. To study the problem in more detail, alternative solvable gravity models can be used, which may help understanding the quantum aspects of Einstein gravitation. Here a gravity model in (1 + 1) space-time dimensions with a dilaton field is adapted, which is further explained in appendix A. The solution with a collapsing null-matter-shell in such a model, also known as the CGHS[3] black hole, has been found at the classical level and can also be formally quantised canonically [12, 13]. The conformal diagram of the classical solution is shown in fig. 2.2.

At the next-to-leading order in the semi-classical approximation scheme, the matter field can be separated from the gravity and the dilaton, and a quantum theory of fields in curved space-time can be *derived* in terms of a functional Schrödinger equation

$$i\frac{\partial \chi[f]}{\partial t} = \int dx \frac{1}{2}\left\{-\frac{\delta^2}{\delta f^2} + \left(\frac{\partial f}{\partial x}\right)^2\right\}\chi[f], \qquad (2.2.1)$$

where $f(x)$ is the classical matter field promoted to an operator, and $\chi[f]$ the corresponding matter wave-functional. Before the collapse of the ingoing null-matter-shell where the region is called *linear dilaton vacuum*, a 'vacuum' state can be found, whose wave functional reads

$$\chi_0[f] \propto \exp\left\{-\frac{1}{2}\int_0^{+\infty} dk\, k\, f(k)^2\right\}, \qquad (2.2.2)$$

---

[3] Introduced by Callan, Giddings, Harvey, and Strominger in [12].

where $f(k)$ is the Fourier sine transform of $f(x)$. Equation (2.2.2) is nothing else but a generalisation of the ground-state wave function of a simple harmonic oscillator, see appendix B.

After the collapse where the region is called *black hole*, the spacial slice is shifted, and so are the Fourier modes of the matter field. It can be shown that

$$f(k) = \int_{-\infty}^{+\infty} \mathrm{d}l\, \alpha(k;l) g(l) \qquad k > 0, \tag{2.2.3}$$

where $g(l)$ is the Fourier transform of matter field after the ingoing shock wave. The Bogolyubov-type coefficient $\alpha(k;l)$ has been computed, substituting which into eq. (2.2.2) yields

$$\chi_b[g] \propto \exp\left\{-\int_{-\infty}^{+\infty} \mathrm{d}p\, p \coth\left(\frac{\pi p}{2\lambda}\right) |g(p)|^2\right\} \tag{2.2.4}$$

in the black hole region, which is a squeezed-state wave functional [4] and obviously different from a ground-state wave functional. The particle-number expectation of eq. (2.2.4) reads

$$\langle n(k) \rangle_{\chi_b} = \left(\mathrm{e}^{2\pi|k|/\lambda} - 1\right)^{-1}, \tag{2.2.5}$$

leading to a Hawking-like *black-body* temperature

$$T_{\mathrm{HD}} := \lambda/2\pi. \tag{2.2.6}$$

Here it is the cosmological constant $\lambda$ which takes the place of surface gravity $\kappa$ in eq. (2.1.6).

## 2.3 Discussion

Though used throughout this work, the density operators can be mathematically defined only when they are *trace-class* for which a trace may be defined, and this is not proved for any of the cases concerned. Unfortunately, the density operators are often not trace-class, especially in quantum field theory. A rigorous approach to deal with such thermal states can be the Kubo–Martin–Schwinger state [14–16] defined in algebraic quantum theory, which is concisely introduced in [17, ch. 3].

Alert readers may be concerned about the result eq. (2.2.4), which is derived in dilaton gravity model, not from Einstein gravitation. The solution is believed to be physically relevant, not only because it gives a same Hawking-like temperature for the CGHS black hole. It has also been shown that the wave functional of a massless, neutral scalar field in the Unruh effect yields exactly the same particle-number fluctuation and Hawking temperature [4, 18]. People have argued and widely believed that the Unruh effect is closely connected to Hawking radiation in Einstein gravity, especially in the space-time region near the horizon (e.g. [19]). It is therefore reasonable to believe that the aforementioned results from the CGHS model give hints about reality.

# Correlator of Field Strength

Having revealed the discrepancy in the kinematic description of Hawking radiation, namely as either a pure or a thermal and mixed state, the author will show the difference in a physically relevant form. Recall that the particle-number expectations of the pure and mixed states are the same, i.e. the diagonal elements of the density operators in the particle-number basis are the same. Hence one seeks operators revealing the off-diagonal elements of the density operator.

## 3.1 Correlator and Fluctuation of Fourier Modes

A natural candidate to reveal the off-diagonal elements of the density operators is the correlator. Bearing in mind the correlator of generalised Gaussian wave functions (see eq. (B.2.3)), the correlator of the wave functional eq. (2.2.4) can be read off as

$$\langle g^\dagger(p_1)g(p_2)\rangle_{\chi_b} = \frac{1}{2}\Big(\langle g_\Re(p_1)g_\Re(p_2)\rangle_{\chi_b} + \langle g_\Im(p_1)g_\Im(p_2)\rangle_{\chi_b}\Big) = \frac{1}{2}\frac{\tanh\frac{\pi p_1}{2\lambda}}{p_1}\delta(p_1-p_2)$$
$$\propto \frac{1}{8T_{\text{HD}}}\frac{\tanh(q/4)}{q/4} = \frac{1}{8T_{\text{HD}}}(1+\text{O}(q)) \qquad (3.1.1)$$

by substituting

$$g(p) = 2^{-1/2}(g_\Re(p) + \mathring{\imath}g_\Im(p)), \qquad (3.1.2)$$

where $2^{-1/2}g_\Re(p)$ and $2^{-1/2}g_\Im(p)$ are the real and imaginary part of $g(p)$, respectively. In the last line of eq. (3.1.1), the delta function is ignored, and the dimensionless parameter

$$q := p_1/T_{\text{HD}} \equiv 2\pi p_1/\lambda \qquad (3.1.3)$$

has been used. By the same argument, one also solves the correlator

$$\langle g^\dagger(p_1)g(p_2)\rangle_{\text{vac}} = \frac{1}{2|p_1|}\delta(p_1-p_2) \propto \frac{1}{2T_{\text{HD}}}q^{-1} \qquad (3.1.4)$$

for the vacuum wave functional $\exp\{-\int_{-\infty}^{+\infty} dp\,|p||g(p)|^2\}$. For the thermal state, on the other hand, the correlator follows from eq. (B.2.5), so that

$$\langle g^\dagger(p_1)g(p_2)\rangle_{\text{th}} = \frac{\coth\frac{\pi p_1}{\lambda}}{2p_1}\delta(p_1-p_2)$$
$$\propto \frac{1}{4T_{\text{HD}}}\frac{\coth(q/2)}{q/2} = \frac{1}{4T_{\text{HD}}}\left(\left(\frac{q}{2}\right)^{-2} + \frac{1}{3} + \text{O}(q)\right). \qquad (3.1.5)$$

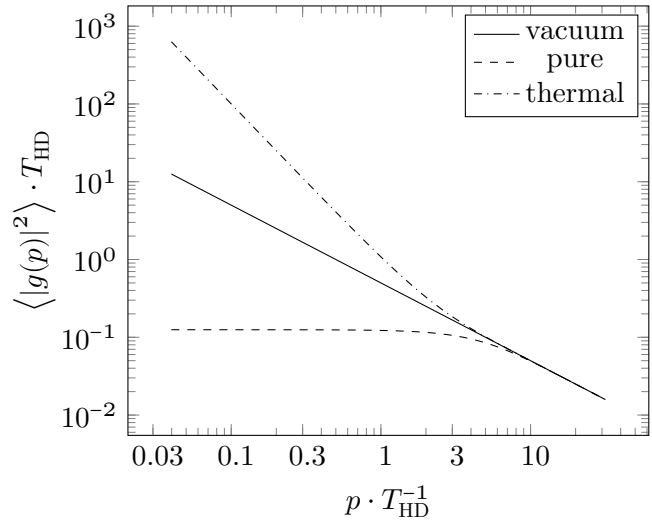

Figure 3.1: Fluctuations of the Fourier modes of the field, plotted in log–log scales. The critical energy scale is the Hawking temperature, below which the various fluctuations depart. The discrepancy between the pure and thermal descriptions is most significant for low-energy modes, while for high energies the fluctuations of them and the vacuum state are practically the same. Moreover, compared with vacuum, the low-energy fluctuation of thermal case is enhanced, so it diverges faster than that of vacuum; for the pure description, however, the fluctuation is suppressed, such that it converges to a constant of order unity and does not diverge any more.

One sees immediately that eqs. (3.1.1), (3.1.4) and (3.1.5) vanish identically for $p_1 \neq p_2$, which is due to the absence of interaction in the scalar field. The remaining diagonal terms have the meaning of *fluctuation* in field strength, because $\langle g \rangle \equiv 0$, so that

$$\left\langle (\Delta g)^2 \right\rangle = \langle g^2 \rangle - \langle g \rangle^2 \equiv \langle g^2 \rangle. \tag{3.1.6}$$

Ignoring the $\delta(0)$ divergence, the fluctuations are plotted in fig. 3.1.

## 3.2 Discussion

It has been noticed that the fluctuation discussed above has been extensively used in cosmology, see e.g. [20]. The fluctuation in the radiation field, however, looks different from that in cosmology and thus remains yet to be explained.

Moreover, the most natural candidate to reveal the off-diagonal elements is the correlator in *real space*, which is just the Fourier transform of the correlator calculated above. Unfortunately, the transformation has yet to be made for the thermal state due to an infra-red divergence.

# Quantification of the Difference

Having shown explicitly the discrepancy between the pure and the thermal descriptions as well as where it is most significant, the author will then quantify the difference. Related definitions of distances in quantum mechanics are introduced in appendix C.

## 4.1 The Canonical State

It has been shown in [21] that a small subsystem in a large, isolated system which is subject to a *constraint*, is expected to be very close to a *canonical state* of it, which reduces to the usual thermodynamic *canonical ensemble* when the total system is under an *energy constraint*.

Denoted by $U$, the large system has all its possible pure states in Hilbert space $\mathcal{H}_U$ which has a finite dimension $d_R = \dim \mathcal{H}_U$. A global constraint $R$ is also imposed, which restricts the physical Hilbert space of $U$ to $\mathcal{H}_R \subseteq \mathcal{H}_U$. The *equiprobable state* of $U$ is

$$\mathcal{E}_R := d_R^{-1} \mathbb{1}_R, \qquad (4.1.1)$$

where $\mathbb{1}_R$ is the identity operator on $\mathcal{H}_R$. An energy constraint, for instance, reads

$$\langle H_U \rangle = E_R, \qquad (4.1.2)$$

so that a state $|\alpha\rangle$ in $\mathcal{H}_R^{(\text{E})}$ satisfies

$$E_R = \langle \alpha \,|\, H_U \,|\, \alpha \rangle = \sum_E |\langle E \,|\, \alpha \rangle|^2 E, \qquad (4.1.3)$$

where $|E\rangle$ is an energy eigenstate of $U$ with eigenvalue $E$.

The subsystem in concern is named $S$, with *all possible* states in the Hilbert space $\mathcal{H}_S$; the rest of $U$ is called the *environment* and denoted by $E$, the pure states of which lie in $\mathcal{H}_E$. One has

$$\mathcal{H}_S \otimes \mathcal{H}_E = \mathcal{H}_U \supseteq \mathcal{H}_R, \qquad H_U = H_S + H_E + H_{\text{int}}, \qquad (4.1.4)$$

where $H_S$, $H_E$ and $H_{\text{int}}$ are the system, environment and interaction Hamiltonian, respectively.

The *canonical state* of $S$ is defined as

$$\Omega_S := \operatorname{tr}_E \mathcal{E}_R, \qquad (4.1.5)$$

where the trace is taken over all the degrees of freedom in the environment. When the energy constraint eq. (4.1.2) is used, it can be shown that $\Omega_S^{(E)}$ is a canonical ensemble[1]

$$\Omega_S^{(E)} \propto e^{-H_S/T_m} = \sum_{E_S} e^{-E_S/T_m} |E_S\rangle \langle E_S|, \qquad (4.1.6)$$

where $T_m$ can be identified with a temperature, and $|E_S\rangle$ is an eigenstate of the system with eigenvalue $E$. In this case it reduces to the traditional thermodynamic statistical physics.

The environment also has an *effective* dimension

$$d_E^{\text{eff}} := (\operatorname{tr} \Omega_E^2)^{-1} \geq d_R/d_S, \qquad (4.1.7)$$

where $\Omega_E = \operatorname{tr}_S \mathcal{E}_R$. When no constraint is enforced, so that $\mathcal{H}_S \otimes \mathcal{H}_E \equiv \mathcal{H}_U = \mathcal{H}_R$, eq. (4.1.7) reduces to

$$d_E^{\text{eff}} = d_R/d_S = d_E. \qquad (4.1.8)$$

Detailed discussions about $d_E^{\text{eff}}$ can be found in [21].

Equipped with all the definitions above, one picks an arbitrary pure state $|\phi\rangle \in \mathcal{H}_R$ and denote the *reduced state* of $S$ by

$$\rho_S(\phi) = \operatorname{tr}_E |\phi\rangle \langle \phi| \qquad (4.1.9)$$

Then a lemma states that the average *trace distance*[2] between $\rho_S$ and $\Omega_S$ is very small in terms of the ratio between $d_S$ and $d_S/d_E^{\text{eff}}$, i.e.

$$\langle T(\rho_S(\phi), \Omega_S) \rangle \leq \frac{1}{2} \sqrt{\frac{d_S}{d_E^{\text{eff}}}}. \qquad (4.1.10)$$

In a typical division where $d_S/d_E$ is small, this average distance will also be tiny.

More over, the main theorem asserts that those $\rho_S$'s which are close to $\Omega_S$ dominate; the probability of a large deviation is exponentially small with respect to the deviation. For an arbitrary $\epsilon > 0$, the theorem states that

$$\frac{V[\{|\phi\rangle \in \mathcal{H}_R | T(\rho_S(\phi), \Omega_S) \geq \eta\}]}{V[\{|\phi\rangle \in \mathcal{H}_R\}]} \leq \eta', \qquad (4.1.11)$$

where

$$\eta = \epsilon + \frac{1}{2}\sqrt{\frac{d_S}{d_E^{\text{eff}}}}; \qquad \eta' = 4\exp(-C d_R \epsilon^2), \quad C = \frac{2}{9\pi^3}. \qquad (4.1.12)$$

To understand the theorem, first note that the left-hand side of eq. (4.1.11) is a probability measure. More over, one can choose $\epsilon = d_R^{-1/3}$ as well, so that

$$\eta = \epsilon + \frac{1}{2}\sqrt{\frac{d_S}{d_E^{\text{eff}}}} \gtrsim d_R^{-1/3}; \qquad \eta' = 4\exp(-C d_R \epsilon^2) = 4\exp\bigl(-C d_R^{+1/3}\bigr), \qquad (4.1.13)$$

where $d_E^{\text{eff}} \gg d_S$ is also assumed. In this case, the probability of the deviation greater than $d_R^{-1/3}$ is smaller than an *exponential* of $d_R^{+1/3}$.

---

[1] A derivation of eq. (4.1.6) in classical statistical physics can be found in [22, sec. 28].
[2] See appendix C.1.

Further explanations and proofs of the lemma and the theorem can be found in [21, 23].

## 4.2 Distance between the Pure and Mixed States

To apply the aforementioned formalism to the Hawking radiation, one recognises the universe in the CGHS model as the total isolated system $U$, whereas the gravitational degrees of freedom as $E$, and the system in concern $S$ is the radiation field. It has been shown in [4, 13] that the reduced state of the radiation field can indeed be *thermal* in certain *decoherence* schemes, while in the collapsing case discussed in section 2.2 and appendix A, the radiation field remains pure.

In this section, the trace distance between the pure and thermal descriptions will be evaluated. Due to technical difficulties, the distance has not been able to be derived exactly. Instead, a lower and an upper bound have been set to the distance by *Fuchs-van de Graaf inequality* in eq. (C.2.4), where only the calculable *Fidelity* (see appendix C.2) is needed.

Note that the wave functional of the Hawking radiation field eq. (2.2.4) is, roughly speaking, the superposition of quantum-mechanical wave functions per Fourier mode,

$$\chi_b[g] \sim \sum_p \chi_b^{(p)}(g_p) = \sum_p \left\langle g_p \, \big| \, \chi_b^{(p)} \right\rangle, \tag{4.2.1}$$

where $g_p := g(p)$ is the Fourier transform of the field $g$ evaluated at $p$. On the other hand, the thermal state of the free field can also be sloppily written as the product of the quantum-mechanical density operators per mode,

$$\rho_{\text{th}} \sim \bigotimes_p \rho_{\text{th}}^{(p)}. \tag{4.2.2}$$

By eq. (C.2.3), the fidelity of $\chi_b[g]$ and $\rho_{\text{th}}$ can be reduced to the product of the fidelity per mode, because

$$\langle \chi_b \, | \, \rho_{\text{th}} \, | \, \chi_b \rangle \sim \prod_p \left\langle \chi_b^{(p)} \, \big| \, \rho_{\text{th}}^{(p)} \, \big| \, \chi_b^{(p)} \right\rangle. \tag{4.2.3}$$

Since $\left|\chi_b^{(p)}\right\rangle$ is just a quantum-mechanical general Gaussian state (see appendix B.1), the result in eq. (C.2.8) can be adapted by substituting

$$\Omega = |p|, \qquad \omega = p \coth \frac{\pi p}{2\lambda} \quad \text{and} \quad T = T_{\text{HD}} \equiv \frac{\lambda}{2\pi}, \tag{4.2.4}$$

yielding the *fidelity per mode*

$$F^{(p)} = \frac{\sqrt{u-1}}{\sqrt[4]{u^2+u+1}}, \qquad u := e^q \equiv e^{|p|/T_{\text{HD}}}, \tag{4.2.5}$$

so that the trace distance per Fourier mode can be evaluated, see fig. 4.1.

To approach the trace distance for the wave functional and the total thermal state, one needs to deal with the product with continuous index in eq. (4.2.3). A popular way to go to the 'continuous limit' in *summation* is

$$\sum_p g(p) \to \frac{1}{2\pi\Lambda} \int dp \, g(p), \tag{4.2.6}$$

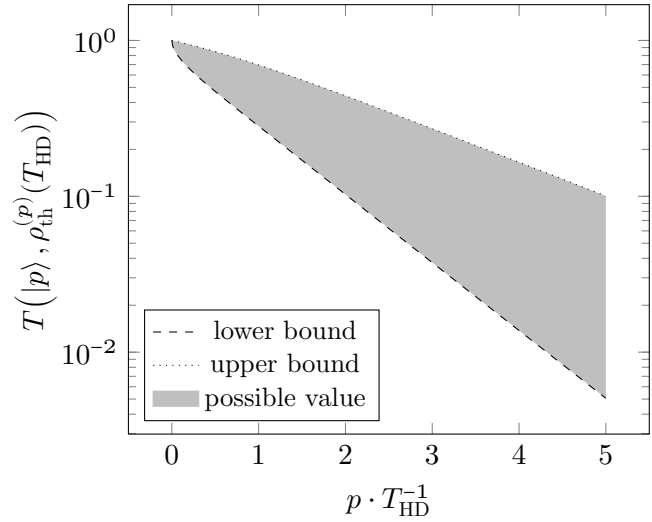

Figure 4.1: Possible value of the trace distance between mode wave functions in eq. (4.2.1) and the corresponding thermal density operators. One sees that the difference becomes exponentially small with respect to $p/T_{\text{HD}}$, which suggests it would be difficult to distinguish the pure and thermal descriptions by detecting the high-energy modes in the Hawking radiation, confirming the results in section 3.1.

where $\Lambda$ has the same dimension as $p$ in order to fix the dimension; in the box-normalisation scheme, for example, the corresponding $\Lambda$ would be proportional to the volume of the box $V$. A similar method may be used to normalise the product, namely

$$\prod_p f(p) = \exp\left\{\sum_p \ln f(p)\right\} \to \exp\left\{\frac{1}{2\pi\Lambda}\int \dd p \ln f(p)\right\}. \tag{4.2.7}$$

Note that the wave functional eq. (2.2.4), which has always been dealt with, can also be seen as being normalised by the method. One thus derives

$$F = \exp\left\{\frac{2}{2\pi\Lambda}\int_0^{+\infty} \dd p \ln F^{(p)}\right\} = \exp\left(-\frac{\pi}{9}\frac{T_{\text{HD}}}{\Lambda}\right), \tag{4.2.8}$$

which is shown in fig. 4.2.

## 4.3 Discussion

Strictly speaking, the formalism in section 4.1 is *not* applicable to the radiation field considered here, because they were proved for *finite dimensional systems*, while the cases in this work are all *infinite-dimensional*. However, since a regularised result can be obtained, it is believable that a mathematically rigorous approach exists, which will justify the result, similar to the renormalisation procedure in quantum field theory.

Though the asymptotic behaviour of the trace distance with respect to Hawking temperature has clearly been revealed by setting bounds to it, it is still appealing to calculate its *exact value*. One possible approach is by using the results in [24], where the eigenfunctions and eigenvalues of a general Gaussian density operator have been explicitly solved.

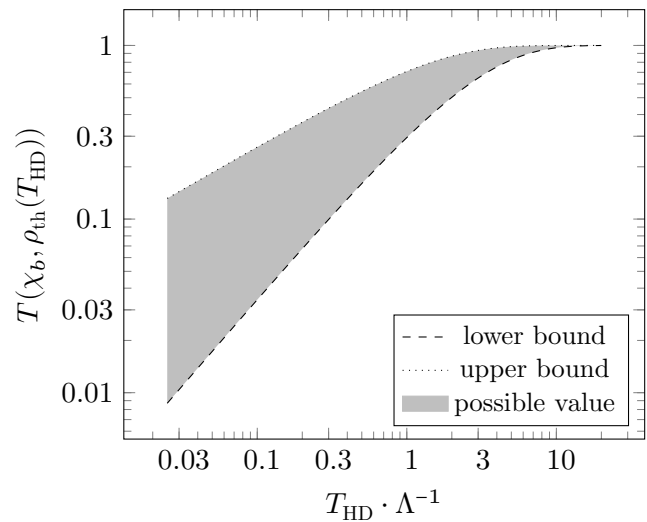

Figure 4.2: Possible value of the trace distance between the pure and thermal descriptions of the Hawking radiation field. One sees that the difference between the two cases is positively (negatively) correlated with the Hawking temperature temperature (mass of black hole).

The evaluation of the trace distance was motivated by Hsu and Reeb in [5], where operational meaning of the distance was also discussed in terms of toy models, which is unfortunately not applicable to the field system. In future works, it is expected that the distance between the pure and thermal states will be understood in terms of *experiments* or *observations*, where the essence of physical science lies.

# Summary and Outlook

In this work, the discrepancy of the pure-state and thermal descriptions of Hawking radiation field has been revealed, based on the CGHS gravity model. It has been shown that in fluctuations of Fourier modes, the difference is significant for low-energy excitations while vanishing for high-energy ones, which fits the result that the Hawking-radiation particles are mostly low-energetic, compared with the Hawking temperature. Moreover, motivated by the new foundation of statistical physics, the discrepancy between the two descriptions has been quantified and proved to be large for a low-mass black hole, which is expected to show more traces of quantum gravity.

The CGHS model used throughout is only one of the solvable alternative gravity models. One may further attempt other popular choices, for example, the *Bañados–Teitelboim–Zanelli model* [25] in (2 + 1) dimensions, and compare the relevant results.

Equipped with the results in the thesis, one expects deeper understandings about the nature of the black-hole entropy and the final fate of the black-hole evaporation, which have been in the spotlight since the advent of Hawking radiation.

In future works, the decoherence formalism will also be taken into account, in which a thermalised reduced state of the radiation field can be obtained explicitly.

# $(1 + 1)$-dimensional Dilaton Gravity

Since quantum field theories in $(3 + 1)$-dimensional Einstein gravitation are difficult to solve, one may turn to alternative solvable gravity models to get some hints for the physics in reality. The $(1 + 1)$-dimensional dilaton gravity, or the Callan–Giddings–Harvey–Strominger (CGHS) model, is such a candidate, which is used in the thesis and has been extensively studied in the literature, see for instance [12, 13, 26]. In this chapter a very brief review of the model is provided, mainly based on [13].

The action for the model, with $N$ massless scalar fields minimally coupled, reads

$$S := \int \mathrm{d}^2 x \sqrt{-\overline{g}} \left\{ \frac{\mathrm{e}^{-2\overline{\phi}}}{G} \left[ \overline{R} + 4(\nabla \overline{\phi})^2 + 4\lambda^2 \right] - \frac{1}{2} \sum_{i=1}^{N} (\nabla f_i)^2 \right\}, \qquad (A.0.1)$$

where $f_i$'s are the neutral scalar matter fields, $\overline{\phi}$ is the dilaton field, $G$ the dimensionless Newton constant, and $\lambda > 0$ the cosmological constant. The dilaton field is essential in two dimensions, because there is only one independent component in the Riemann curvature tensor, hence a pure $(1 + 1)$-dimensional Einstein theory shall be trivial. Transforming with $\phi = \mathrm{e}^{-2\overline{\phi}}$ and $g_{\alpha\beta} = \mathrm{e}^{-2\overline{\phi}} \overline{g}_{\alpha\beta}$ eliminates the kinetic term for the dilaton, yielding

$$S = \int \mathrm{d}x \, \mathrm{d}t \sqrt{-g} \left\{ \frac{1}{G} \left[ R\phi + 4\lambda^2 \right] - \frac{1}{2} (\nabla f)^2 \right\}, \qquad (A.0.2)$$

where only one matter field is considered for simplicity, and an ADM-like[1] parametrisation of the metric

$$\mathrm{d}s^2 = \mathrm{e}^{2\rho} \left[ -\sigma^2 \mathrm{d}t^2 + (\mathrm{d}x + \xi \, \mathrm{d}t)^2 \right] \qquad (A.0.3)$$

is assumed, in which $(\sigma, \xi)$ are the lapse and shift functions. The action in eq. (A.0.2) has a classical solution describing a collapsing null-matter shell, which resembles the solution of a spherically collapsing body in the $(3 + 1)$-dimensional Einstein case. The corresponding conformal diagram is plotted in fig. 2.2.

To apply the canonical quantisation scheme, the action in eq. (A.0.2) is to be recast in the Hamiltonian formalism. However, the Legendre transformation of the field momenta proves to be singular. This means that the momenta, as functions of the positions and velocities, cannot be inverted to express the corresponding velocities as functions of momenta and positions, so

---

[1] Arnowitt–Deser–Misner, see [27].

*Appendix A  (1 + 1)-dimensional Dilaton Gravity*

that the standard algorithm to obtain the Hamiltonian

$$H := \left[\frac{\partial L}{\partial \dot{X}_i}\dot{X}_i - L\right]_{\dot{X}=\dot{X}(P,X)} \tag{A.0.4}$$

does not apply, where $(X, \dot{X}, P)$ are the positions, velocities and momenta, respectively.

The systems, of which the Legendre transformation is singular, are called *constrained systems*. Other examples include a relativistic point particle in the covariant form, Yang–Mills theories and string theories. For such systems, the usual quantisation scheme and the Schrödinger equation do not apply directly. Instead, one has to identify the constraints in the system and apply Dirac's quantisation rules [28, 29]. The result is that the quantum wave functional describing the CGHS model is *constrained* by

$$0 = \mathcal{H}_\parallel \Psi[\rho, \phi, f] := \left(\frac{\mathrm{d}\rho}{\mathrm{d}x}\frac{\delta}{\delta\rho} - \frac{\mathrm{d}}{\mathrm{d}x}\frac{\delta}{\delta\rho} + \frac{\mathrm{d}\phi}{\mathrm{d}x}\frac{\delta}{\delta\phi} + \frac{\mathrm{d}f}{\mathrm{d}x}\frac{\delta}{\delta f}\right)\Psi[\rho, \phi, f], \tag{A.0.5}$$

$$0 = \mathcal{H}_\perp \Psi[\rho, \phi, f] := \left(\frac{G}{2}\frac{\delta^2}{\delta\rho\,\delta\phi} - \frac{1}{2}\frac{\delta^2}{\delta f^2} + \frac{1}{2G}V_G + V_M\right)\Psi[\rho, \phi, f], \tag{A.0.6}$$

where

$$V_G := 4\left(\frac{\mathrm{d}^2\phi}{\mathrm{d}x^2} - \frac{\mathrm{d}\phi}{\mathrm{d}x}\frac{\mathrm{d}\rho}{\mathrm{d}x} - 2\lambda^2 \mathrm{e}^{2\rho}\right), \qquad V_M := \frac{1}{2}\left(\frac{\mathrm{d}f}{\mathrm{d}x}\right)^2. \tag{A.0.7}$$

Equations (A.0.5) and (A.0.6) are the *Wheeler–DeWitt equations* for the CGHS model, which play the role of the usual Schrödinger equation for the whole system.

In the next step, a semi-classical approximation (see also [30, sec. 5.4]) of the Born–Oppenheimer type is applied to $\Psi$ by expanding the exponent as

$$\Psi[\rho, \phi, f] = \mathrm{e}^{\mathring{\mathrm{i}}(G^{-1}S_0 + S_1 + GS_2 + \ldots)}. \tag{A.0.8}$$

Inserting this expression into eqs. (A.0.5) and (A.0.6), one finds that at order $G^0$, variables can be separated by setting

$$\mathrm{e}^{\mathring{\mathrm{i}}S_1} := D^{-1}[\rho, \phi]\chi[\rho, \phi, f]. \tag{A.0.9}$$

Inserting the leading and next-to-leading order terms into eqs. (A.0.5) and (A.0.6) yields

$$\mathring{\mathrm{i}}\left(\frac{\partial\rho}{\partial t}\frac{\delta\chi}{\delta\rho} + \frac{\partial\phi}{\partial t}\frac{\delta\chi}{\delta\phi}\right) = \frac{1}{2}\left\{-\frac{\delta^2}{\delta f^2} + \left(\frac{\partial f}{\partial x}\right)^2\right\}\chi, \tag{A.0.10}$$

integrating of which gives the functional Schrödinger equation for the matter field (2.2.1).

# Harmonic Oscillators

## B.1 Single Harmonic Oscillator

A single harmonic oscillator is described by the Hamiltonian

$$H = \frac{1}{2}(\pi^2 + \Omega^2 f^2), \tag{B.1.1}$$

where $(\pi, f)$ are conjugate momentum and position, respectively. Here a unit system similar to the case in field theory is adapted. Canonical quantisation uses the annihilation and creation operators

$$a^- := \frac{1}{\sqrt{2}}\left(\sqrt{\Omega} f + \frac{\mathring{\imath}}{\sqrt{\Omega}}\pi\right), \qquad a^+ := \frac{1}{\sqrt{2}}\left(\sqrt{\Omega} f - \frac{\mathring{\imath}}{\sqrt{\Omega}}\pi\right) \equiv (a^-)^\dagger; \tag{B.1.2}$$

inverse expressions read

$$f = \frac{a^+ + a^-}{\sqrt{2\Omega}}, \qquad \pi = \mathring{\imath}\sqrt{\frac{\Omega}{2}}(a^+ - a^-). \tag{B.1.3}$$

The ground state and the normalised $n$th excitation are defined by

$$a^- |0\rangle := 0, \qquad |n\rangle := \frac{1}{\sqrt{n!}}(a^+)^n |0\rangle, \qquad n \in \mathbb{Z}_+, \tag{B.1.4}$$

the wave functions of which are

$$\langle f | n \rangle = \frac{1}{\sqrt{2^n n!}} e^{-\Omega f^2/2} H_n\left(\sqrt{\Omega} f\right), \qquad n \in \mathbb{Z}, \tag{B.1.5}$$

where $H_n(x)$ is the $n$th Hermite polynomial. The normalising measure

$$\mathrm{d}\mu(f) := \sqrt{\frac{\Omega}{\pi}}\, \mathrm{d}f \tag{B.1.6}$$

is used throughout, so that the completeness relation holds,

$$\int \mathrm{d}\mu(f)\, \langle \alpha | f \rangle \langle f | \beta \rangle \equiv \langle \alpha | \beta \rangle. \tag{B.1.7}$$

In the present work, a general Gaussian state $|\omega\rangle$ has also been considered, the wave function of which reads

$$\langle f | \omega \rangle = \left(\frac{\mathfrak{R}\omega}{\Omega}\right)^{1/4} \exp\left(-\frac{\omega}{2} f^2\right), \qquad \mathfrak{R}\omega > 0 \tag{B.1.8}$$

Evaluating the expectation of $f^2$ yields

$$\langle f^2 \rangle_\omega := \langle \omega | f^2 | \omega \rangle = (2\mathfrak{R}\omega)^{-1}. \tag{B.1.9}$$

$|\omega\rangle$ can also be expressed in terms of energy eigenstates,

$$\begin{aligned}
\langle n | \omega \rangle &= \int d\mu(f) \, \langle n | f \rangle \langle f | \omega \rangle \\
&= \left(\frac{\Omega\mathfrak{R}\omega}{\pi^2}\right)^{1/4} \frac{1}{\sqrt{2^n n!}} \int_{-\infty}^{+\infty} df \exp\left\{-\frac{1}{2}(\Omega+\omega)f^2\right\} H_n\left(\sqrt{\Omega}f\right) \\
&= \begin{cases} (\Omega\mathfrak{R}(\omega))^{\frac{1}{4}} \dfrac{2^{\frac{1}{2}-m}\sqrt{(2m)!}}{m!} \dfrac{(\Omega-\omega)^m}{(\omega+\Omega)^{m+\frac{1}{2}}} & n = 2m, \\ 0 & n = 2m+1, \end{cases}
\end{aligned} \tag{B.1.10}$$

thanks to [31].

A thermal state of the oscillator at temperature $T$ can be described by the density operator

$$\rho := \frac{1}{Z}\exp\left\{-\frac{H}{T}\right\} = \frac{1}{Z}\sum_{n=0}^{+\infty}\exp\left\{-\frac{\Omega}{T}\left(n+\frac{1}{2}\right)\right\}|n\rangle\langle n|, \tag{B.1.11}$$

where the partition function is

$$Z := \operatorname{tr} e^{-H/T} = \frac{1}{2}\operatorname{csch}\frac{\Omega}{2T}. \tag{B.1.12}$$

One also obtains

$$\begin{aligned}
\langle f^2 \rangle_\rho &= \frac{1}{2\Omega}\frac{1}{Z}\sum_{n=0}^{+\infty}\exp\left\{-\frac{\Omega}{T}\left(n+\frac{1}{2}\right)\right\}\langle n|(a^+ + a^-)^2|n\rangle \\
&= \frac{1}{2\Omega}\frac{1}{Z}\sum_{n=0}^{+\infty}\exp\left\{-\frac{\Omega}{T}\left(n+\frac{1}{2}\right)\right\}\langle n|(2a^+a^- + 1)|n\rangle = \frac{1}{4T}\frac{\coth\frac{\Omega}{2T}}{\frac{\Omega}{2T}}.
\end{aligned} \tag{B.1.13}$$

## B.2 Multiple Harmonic Oscillators

Multiple harmonic oscillators are described by the Hamiltonian

$$H = \sum_i H_i, \qquad H_i = \frac{1}{2}(\pi_i^2 + \Omega^2 f_i^2), \tag{B.2.1}$$

where $(\pi_i, f_i)$ are conjugate momentum and position of the $i$th oscillator.

When the general Gaussian state

$$\langle \{f\} | [\omega] \rangle = \left(\det\frac{\omega}{\Omega}\right)^{1/4}\exp\left(-\frac{1}{2}f_i\omega_{ij}f_j\right) \tag{B.2.2}$$

is considered, where $\omega_{ij}$ is a real symmetric positive-definite matrix, the two-point correlator becomes

$$\langle [\omega] | f_i f_j | [\omega] \rangle = \left[(2\omega)^{-1}\right]_{ij}, \tag{B.2.3}$$

which is a generalisation of eq. (B.1.9).

A thermal state at temperature $T$ can be described by the density operator

$$\rho = \bigotimes_i \rho_i, \qquad \rho_i = \frac{1}{Z}\exp\left(-\frac{H_i}{T}\right). \tag{B.2.4}$$

Computing the two-point correlator of the state, one finds

$$\left\langle f_i f_j \right\rangle_\rho = \frac{\coth\frac{\Omega_i}{2T}}{2\Omega_i}\delta_{ij} = \frac{1}{4T}\frac{\coth\frac{\Omega_i}{2T}}{\frac{\Omega_i}{2T}}\delta_{ij} \tag{B.2.5}$$

from eq. (B.1.13).

# Distances in Quantum Theory

This chapter follows mostly [32]. Some conventions follow those in [33].

## C.1 Trace Distance

To begin with, define the *trace norm* or $l_1$-*norm* of an Hermitian operator $M$ as

$$\|M\|_1 := \operatorname{tr} \sqrt{M^\dagger M}. \tag{C.1.1}$$

When the spectral decomposition $M = \sum_i \mu_i \ket{i}\bra{i}$ exists, the trace norm reads

$$\|M\|_1 \equiv \sum_i |\mu_i|, \tag{C.1.2}$$

so the name $l_1$-norm comes. It is positive definite and homogeneous; the triangle inequality also holds. Thus it can be used to define the *trace distance* between Hermitian operators $M$ and $N$ as

$$T(M, N) := \frac{1}{2}\|M - N\|_1 \equiv \frac{1}{2} \operatorname{tr} \sqrt{(M-N)^\dagger(M-N)}. \tag{C.1.3}$$

Now consider density operators $\rho$ and $\sigma$ only. Since $\|\rho\|_1 \equiv 1$, one sees

$$0 \leq T(\rho, \sigma) \leq 1, \tag{C.1.4}$$

followed from positive definiteness and triangle inequality.

The following lemma helps constructing a physical interpretation of the distance. Let the Hermitian operator $\Lambda$ be such that all its eigenvalues lies within $[0, 1]$, then

$$T(\rho, \sigma) = \max_{0 \leq \Lambda \leq \mathbb{1}} \operatorname{tr}\{\Lambda(\rho - \sigma)\}. \tag{C.1.5}$$

To understand this, take $\Lambda \equiv \ket{\alpha}\bra{\alpha}$, where $\ket{\alpha}$ is the eigenket of some observable A, with eigenvalue $\alpha$. Then $\operatorname{tr}\{\Lambda\rho\}$ tells the probability of measuring A which gives the result $\alpha$. Therefore $\operatorname{tr}\{\Lambda(\rho - \sigma)\}$ gives the difference of the probability above for $\rho$ and $\sigma$, and $T(\rho, \sigma)$ is the maximal value of the difference above.

Though the trace distance is used in formulating chapter 4, it is rather formidable to compute due to the operatorial square root in eq. (C.1.3). Therefore the author seeks other ways to evaluate the quantity.

## C.2 Fidelity

Fidelity is another means to compare two quantum states. The simplest case of fidelity is that of two pure states,
$$F(|\alpha\rangle, |\beta\rangle) := |\langle \alpha | \beta \rangle|. \tag{C.2.1}$$

One sees that it is just the modulus of the transition amplitude, a measure of *faithfulness*. For more general cases, fidelity is defined as

$$F(|\alpha\rangle, \sigma) := \sqrt{\langle \alpha | \sigma | \alpha \rangle}, \tag{C.2.2}$$

$$F(\rho, \sigma) := \operatorname{tr} \sqrt{\rho^{\frac{1}{2}} \sigma \rho^{\frac{1}{2}}}. \tag{C.2.3}$$

In applications in this work, fidelity is solvable because only eq. (C.2.3) is needed.

An important property of fidelity is that it follows the *Fuchs–van de Graaf inequality* [34])

$$1 - F(\rho, \sigma) \leq T(\rho, \sigma) \leq \sqrt{1 - F^2(\rho, \sigma)}. \tag{C.2.4}$$

In this work, the trace distances are evaluated by computing the fidelities exactly and inserting them to eq. (C.2.4).

Consider a single harmonic oscillator with intrinsic frequency $\Omega$, a thermalised state of it at temperature $T$ described by a density operator $\rho(T)$, as well as a generalised Gaussian state $|\omega\rangle$ (see appendix B.1). In the following the fidelity of them will be calculated.

Since $\rho(T)$ is diagonal in the energy-eigenstate basis $|n\rangle$, one has

$$\langle \omega | \rho(T) | \omega \rangle = \sum_{n=0}^{+\infty} \langle \omega | n \rangle \langle n | \rho(T) | n \rangle \langle n | \omega \rangle, \tag{C.2.5}$$

in which

$$\langle n | \omega \rangle = \begin{cases} (\Omega \Re \omega)^{\frac{1}{4}} \dfrac{2^{\frac{1}{2}-m}\sqrt{(2m)!}}{m!} \dfrac{(\Omega - \omega)^m}{(\omega + \Omega)^{m+\frac{1}{2}}} & n = 2m, \\ 0 & n = 2m + 1, \end{cases} \tag{B.1.10 revisited}$$

$$\langle n | \rho | n \rangle = \frac{1}{Z} \exp\left\{-\left(n + \frac{1}{2}\right)\frac{\Omega}{T}\right\} \equiv (\mathbb{e}^{\Omega/T} - 1)\exp\left\{-(n+1)\frac{\Omega}{T}\right\}. \tag{C.2.6}$$

Inserting eqs. (B.1.10) and (C.2.6) into eq. (C.2.5), one finds that each term in the summation is of the form $b \cdot \binom{2m}{m} a^{2m}$ where $a$ and $b$ are expressions of $\Omega$, $\omega$ and $T$, which can be computed by using

$$\arcsin z = \sum_{n=0}^{+\infty} \binom{2n}{n} \frac{z^{2n+1}}{4^n(2n+1)} \tag{C.2.7}$$

and taking derivative with respect to $z$ on both sides. Therefore the fidelity of $\rho(T)$ and $|\omega\rangle$ is computed to be

$$F(|\omega\rangle, \rho(T)) = \sqrt{2} \sqrt[4]{\frac{\Omega \Re \omega \left(\mathbb{e}^{\Omega/T} - 1\right)^2}{(\Omega - \Re \omega)^2 + \Im \omega^2 - \mathbb{e}^{2\Omega/T}((\Omega + \Re \omega)^2 + \Im \omega^2)}}, \tag{C.2.8}$$

which can be used to set a bound on the trace distance by eq. (C.2.4).

# List of Figures

# Acknowledgements

I would like to acknowledge the Bonn-Cologne Graduate School for Physics and Astronomy, the Rheinische Friedrich-Wilhelms-Universität Bonn, the Universität zu Köln and my parents, who supported my Master study financially.

I am also grateful to Prof. Dr. Claus Kiefer, Prof. Dr. Hans-Peter Nilles and Prof. Dr. Friedrich W. Hehl, who gave me instructions into Gravity and guided my thesis carefully.

I would also like to thank my fellow group members: Branislav Nikolić, Nick Kwidzinski, Anton Krieger, David Wichmann, Dennis Piontek, Anirudh Gundhi and Matthias Dahlmanns, who helped me developing the ideas as well as commenting the thesis.